\documentclass[
12pt,
reprint,  
showpacs, 
aps, 
prd, 
nofootinbib, 
floatfix, 
amsmath, 
amssymb
]{revtex4-1}
\pdfoutput=1
\usepackage{graphicx}
\usepackage{dcolumn}
\usepackage{bm}
\usepackage{natbib}
\usepackage{float}
\usepackage{color}
\usepackage[
]{hyperref}
\usepackage{mathrsfs}
\usepackage{rotating}
\usepackage{epstopdf}
\usepackage{footnote} 
 \usepackage{subcaption}

\newcommand{\be}{\begin{eqnarray}}
\newcommand{\ee}{\end{eqnarray}}
\newcommand{\nn}{\nonumber}
\newcommand{\nin}{\noindent}

\def \Br{\mathop{\mbox{\normalfont Br}}\nolimits}
\def \Uu{\mathop{\mbox{\normalfont U(1)}}\nolimits}
\def \SUu{\mathop{\mbox{\normalfont SU(2)}}\nolimits}
\def \SUuu{\mathop{\mbox{\normalfont SU(3)}}\nolimits}
\def \Or{\mathop{\mbox{{O}}}\nolimits}
\def \MeV{\mathop{\mbox{\normalfont MeV}}\nolimits}
\def \GeV{\mathop{\mbox{\normalfont GeV}}\nolimits}
\def \TeV{\mathop{\mbox{\normalfont TeV}}\nolimits}
\def \dL{\mathcal{L}}
\def \StMo{\mathop{\mbox{\normalfont SU(3)}}_C \times \SUu_L \times 
\Uu_Y }

\begin{document}
\title{
Exclusions on $Z'$ mass and its 
non-universal couplings in LFV decays
}

\author{J.A. Orduz-Ducuara}
\affiliation{
Departamento de F\'isica, \\
FES-Cuautitl\'an Izcalli, UNAM\\
C.P. 54740, \\
Estado de M\'exico, M\'exico.}

\date{\today}

\begin{abstract}
This letter presents 
a phenomenological analysis for the lepton 
($l_i^{}$)  decaying into  
$ l_j^{} l_k^{} \bar{l}_k^{}, l_j^{} \pi^+ \pi^-$ 
considering  
 family non-universal couplings as source of the lepton flavor-violating (LFV)
currents and a new neutral gauge boson ($Z'$)
as mediator in the flavor-changing. 
The most viable $g_{V}^{f_if_J}$ and 
$g_{A}^{f_if_j}$ couplings are reported  
as long as derive new bounds for the $M_{Z'}^{}$ by using current results from LHC and a phenomenological analysis. 
\end{abstract}

\maketitle

\section{\label{sec:Intro} 
Introduction}

We have different motivations to explore lepton sector, 
taking a new gauge neutral boson as mediator in LFV.
We explore new scenarios with
flavor-changing 
neutral currents (FCNC) 
to obtain 
bounds for the model parameters 
\cite{Demir:2005ti}, and it could imply 
new physics (NP). The current (or future) 
colliders could use the data to discard 
some theoretical models. 

One of the simplest model extends the symmetry group of the standard model (SM). This kind of extension introduces  
an extra symmetry group $\Uu,$ that is labeled 
$\Uu'$ with charge $\lambda.$  
The new symmetry group is: 
$\mathcal{G}={\SUuu}_C \times {\SUu}_L 
\times {\Uu}_Y \times {\Uu}_{\lambda}'.$ 
In this type of extensions, the scalar sector 
would have six degree of freedom: 
four from Higgs doublet, and two from the singlet
\cite{Basso:2010jt}. However we will consider 
that the only difference with respect the SM Lagrangian is 
the introduced potential term 
and the kinetic term for the 
singlet field \cite{Basso:2010pe,Basso:2009gg}.

Nowadays, there are experimental motivations to explore 
new physics scenarios; e.g., 
the recent results from CMS and 
ATLAS collaborations \cite{CMS:2015dxe, Aad:2015pfa} 
and LHCb 
preliminary results for $B^+ \to K^+ \mu^+\mu^-(e^+e^-)$ process
have been an incentive to consider family non-universal (FNU) 
coupling. It will be 
studied by Babar, Belle 
(II) and LHCb \cite{LHCP-2014, Babar-collabo, 
Belle-collabo,CLEO-collabo}.

There are a lot of interesting reports containing physics and phenomenology on $Z'$ in  
different contexts \cite{Langacker:2008yv, Langacker:2009im, 
Leike:1998wr, Erler:1999ub, Abada:2015zea, Bandyopadhyay:2014sma, Crivellin:2015lwa}, 
others that constrain 
the parameters related to $Z'$ \cite{Kundu:1995gr, Erler:2009jh} 
and several letters on  
new neutral gauge bosons and Higgs particle 
\cite{Edelhauser:2014yra, Li:2013ava,Diaz-Cruz:2013kpa, 
Lopez:2013hsa}. 
Some papers about NP with 
universal{\footnote{Flavor violation 
could be suppressed if the charges are family universal  (FU)
\cite{Langacker:2008ip}.}} and  
FNU couplings, which are given 
by the different values 
for the fermion couplings, can be found 
in \cite{Bernabeu:1993ta, Langacker:2000ju, Chiang:2011cv, 
Barger:2009eq, Barger:2009qs, Arhrib:2006sg} 
with interesting phenomenological
results. 
Other papers considering 
FNU couplings for $B-$decays are  
\cite{Barger:2004qc, Chang:2013hba, Li:2011nf, Sirvanli:2013fts, Becirevic:2016zri, Alok:2010ij}.
An increasing number of papers
considering FNU couplings
have appeared recently; e.g., on rare 
semilepton decays \cite{Chang:2011jka},
on leptonic channels including neutrinos 
\cite{La:2013gga, Khan:2016uon, Chiang:2012ww}, and, even 
about G(221) models \cite{Hsieh:2010zr}. 
There also are papers considering a new gauge neutral boson 
coupling to the fermions of the third family 
\cite{Holdom:2008xx, Erler:1999nx}.
In our paper the $\theta'$ (mixing angle $ZZ'$)
parameter appears in the equations, explicitly.
We will fix that parameter respecting the 
precision measurements imposed, 
this is: $\theta' \lesssim 10^{-3}$ \cite{Abreu:1994ria}. Besides we will use some reports to 
explore new models as $E_{6},$ Left-Right and others.

In this letter we will analyze the most representative lepton 
processes in order to constrain the new neutral 
gauge boson mass and its flavor-changing (FC)  non-universal 
couplings. 
Basically, we explore the $g_{V,A}^{f_if_j}$ parameters 
which differs from the current literature where 
$\epsilon_{L,R}$ chiral coupling are taken 
($g_{V,A}^{} = \epsilon_{L}^{} \pm \epsilon_{R}^{}$ 
\cite{Langacker:2008yv}). 
Section \ref{sec:chan-fla-ori} describes the 
FC for this model in fermion sector 
and we show the lagrangian for the model.
Section \ref{sec:leton-processes}, 
presents flavor-violating in 
leptonic sector, and hadron decays.
Finally, in section \ref{sec:discussion} we discuss 
the results and state our conclusions.

\section{\label{sec:chan-fla-ori}
About the model: Lagrangian}

We shall consider a general Lagrangian for 
fermions and new neutral gauge boson which 
is very similar to the standard model. 
In this context, the models could introduce  
new fermions and new SM fermion charges under the new symmetry group.
Then the model obtains FC through: {\it{a)}} the mixing of the SM fermions with 
the new fermions (introduced to avoid anomalies 
\cite{Langacker:2000ju, La:2013gga}), 
{\it{b)}} the SM fermion charges under the extra group can be FNU 
\cite{Barger:2009qs, Cleaver:1998sm, Cleaver:1997jb, 
Masip:1999mk}. 
We are interested in the second method.

In the next subsections we consider different scenarios which have a pedagogical motivation 
and we assume to respect the CKM bounds. 
 
\subsection{
Lagrangian for the family universal model}

In a family universal model, the vertex $\bar{f}f Z'$ 
is proportional to fermion charges:

\be
{\dL^{FU}}  \propto 
\begin{pmatrix}
\bar{f}_1^0 & \bar{f}_2^0 & \bar{f}_3^0
\end{pmatrix}_L
\gamma^\mu
\begin{pmatrix}
Q_L&0&0\\
0&Q_L&0\\
0&0&Q_L
\end{pmatrix}
\begin{pmatrix}
{f}_1^0\\
{f}_2^0\\
{f}_3^0
\end{pmatrix}_L
Z'_\mu
\nn
\ee

\nin
We consider a rotation to the mass eigenstates 
$f_L = V_L f^0_L,$ where 
$f_L^0 = \begin{pmatrix}
{f}_1^0\\
{f}_2^0\\
{f}_3^0
\end{pmatrix}_L$ ($f^0_{}$ means interaction eigenstates) and $V_L$ is a orthogonal transformation matrix; e.g.:
\be
V =
\begin{pmatrix}
 c_{12} c_{13} & 
 c_{13} s_{12} & 
 s_{13} \\
 -c_{23} s_{12}-c_{12} 
 s_{13} s_{23} & c_{12} 
 c_{23} -s_{12} s_{13} 
 s_{23} & c_{13} s_{23} \\
 s_{12} s_{23} -c_{12} 
 c_{23} s_{13} & -c_{23}
 s_{12} s_{13}-c_{12} 
 s_{23} & c_{13} 
 c_{23}\\
\end{pmatrix}
\nn
\ee

\nin
where $c_{ij} = \cos\theta_{ij}$ and 
$s_{ij} = \sin\theta_{ij}.$ Then:

\be
{\dL^{FU}} \propto
\begin{pmatrix}
\bar{f}_1 & \bar{f}_2 & \bar{f}_3
\end{pmatrix}_L
\gamma^\mu Q_L
V_L^\dag
\begin{pmatrix}
1&0&0\\
0&1&0\\
0&0&1
\end{pmatrix}
V_L
\begin{pmatrix}
{f}_1\\
{f}_2\\
{f}_3
\end{pmatrix}_L
Z'_\mu
\nn
\ee

\nin
where  
$Q_L$ is the family universal coupling. 
If the charges are same, matrix 
$\boldsymbol{Q}
\Big( = V_L^\dag ~I~ V_L\Big)$ is diagonal and there is not mixing.

\subsection{
Lagrangian for the family non-universal model}

In this subsection, we present the way the non-universal couplings generate the flavor-change, namely: 

\be
{\dL^{FNU}}  \propto 
\begin{pmatrix}
\bar{f}_1^0 & \bar{f}_2^0 & \bar{f}_3^0
\end{pmatrix}_L
\gamma^\mu
\begin{pmatrix}
Q_L&0&0\\
0&{Q'}_L&0\\
0&0&{Q''}_L
\end{pmatrix}
\begin{pmatrix}
{f}_1^0\\
{f}_2^0\\
{f}_3^0
\end{pmatrix}_L
Z'_\mu.
\nn
\ee

\nin
As before, we obtain,
\be
{\dL^{FNU}}  \!\propto 
\begin{pmatrix}
\bar{f}_1 & \bar{f}_2 & \bar{f}_3
\end{pmatrix}_L
\gamma^\mu
V_L^\dag \!
\begin{pmatrix}
Q_L&0&0\\
0&{Q'}_L&0\\
0&0&{Q''}_L
\end{pmatrix}
\! \! V_L \! \!
\begin{pmatrix}
{f}_1\\
{f}_2\\
{f}_3
\end{pmatrix}_{\!L}\! \!
\! Z'_\mu. 
\nn
\ee
Charges are family non-universal and the matrix $\boldsymbol{Q}$ is not diagonal.  this scenario 
is interesting to explore the FC mediated by new neutral gauge boson.

We will explore models with FNU couplings 
with gauge group given by: 
${\StMo} \times {\Uu'_{\lambda}}.$ 
We labeled  $q_\lambda$ as 
the SM fermion charges  under $\Uu',$ these
are family-depending. 
We could have $Z'$ effects if we suppose 
${M_{Z'}} \lesssim 2 \TeV-3 \TeV,$ and these 
could be detectable in LHC or in 
future colliders{\footnote{Nowadays, there have been interesting results for bosons with masses around sub$-{\GeV}$ as shown in ref. \cite{Heeck:2016xkh}.}}.


\subsection{Sample in the lepton sector}

We recall the Lagrangian for the neutral currents 
sector, which is considered 
\cite{Langacker:1991pg, Amaldi:1987fu, Langacker:2000ju, Andrianov:1998hx}:
\be\label{eq:lagrangian_gral_CN_CS}
-{\dL}_{CN}  & =  &
\sum\limits_{f} g_{1}^{}\bar{\psi}_{f_i}^{} \gamma^\mu 
\Big(
{g}_V^f - 
{g}_A^f \gamma^5\Big) 
\psi_{f_i}^{} Z_\mu^{1} 
 \nn\\
&&
+\sum\limits_{f_i,f_j}{g'_1}\bar{\psi}_{f_i}^{} \gamma^\mu 
\Big(
{g'}_V^{f_if_j} - 
{g'}_A^{f_if_j} \gamma^5\Big)
\psi_{f_j}^{} {Z}_\mu^{2}
\ee

\nin
where $\psi_{f}$ and $\psi_{f_i}^{}$ are the 
weak gauge eigenstates. 
${\widetilde{g}^{f}_{V,A}}$  are the vector and axial-vector 
associated to the $\bar{f}fZ;$ and 
${\widetilde{g'}^{f_if_j}_A}$ those associated 
to couplings $\bar{f}fZ'$ vertex, and we will consider
$g_{1}^{} = \frac{g}{\cos\theta_W}.$ 
$Z^1_\mu\; , Z^2_\mu$ are the gauge eigenstates associated 
to $Z$ and $Z'$ through 
\be
Z^1_\mu  & = & Z_\mu\cos\theta' + {Z'}_\mu \sin\theta',
\nn\\
Z^2_\mu  & =  -&Z_\mu\sin\theta' + {Z'}_\mu \cos\theta',
\nn
\ee

\nin
respectively. We can re-write the eq. 	\eqref{eq:lagrangian_gral_CN_CS} 
as:
\be\label{eq:lagrangiano_gA_gV}
-{\dL}_{CN}  & = &
g_{1}^{}\Big(\cos\theta'
J^\mu_1
+
\frac{g'_1}{g_{1}^{}} \sin\theta'
J^\mu_2
\Big)
Z_\mu \nn
\\
&&
+
g_{1}^{}\Big(-\sin\theta'
J^\mu_1
+
\frac{g'_1}{g_{1}^{}} \cos\theta'
J^\mu_2
\Big)
{Z'}_\mu
\ee

\nin
where $\theta'$ is the mixing angle $Z-Z'$ and 
\be
J^\mu_1 & = & \sum_{f}\bar{\psi}_f \gamma^\mu \big(
{{{g}}^{f}_{V}} - {{{g}}^{f}_A} \gamma^5\big)\psi_f
\nn\\
J^\mu_2 & = & \sum_{f_i,f_j}{g'}\bar{\psi}_{f_i} 
\gamma^\mu \big({{{g'}}^{f_if_j}_V} - 
{{{g'}}^{f_if_j}_A} \gamma^5\big)
\psi_{f_j}
\nn
\ee

\nin
The eq. \eqref{eq:lagrangiano_gA_gV} shows the explicit 
dependence with $g_{V,A}^{f_if_j}.$ Now on we consider 
the neutral current sector for $Z'$, so we will neglect 
the prime label. Now we have a sector 
\begin{itemize}
\item with non-universal cuplings and 
\item proportional to the charges $q_{\lambda}$ 
      through $g_{V,A}^{f_if_j}.$
\end{itemize}

\section{\label{sec:leton-processes}
Phenomenology: lepton and hadron sector
}

We obtained the decay width{\footnote{Similar results can be found in 
\cite{Chiang:2011cv, Yue:2002ja, Langacker:2000ju}}} using FeynCalc package  
\cite{Mertig:1990an}; namely, 
\be\label{eq:W_l-lllb}
\Gamma(l_i\to l_j l_k\bar{l_k}) &&= 
\frac{{g'_1}^4 {m_{l_i}^{}} r_{l_i Z'}^2 \sec^4\theta'}{768 \pi^3} \times
\nn\\&&
\Bigg(
13\mathcal{F}_1 \Big({g_V^{l_k}}^2  + {g_A^{l_k}}^2\Big) 
+
12 \mathcal{F}_2 {g_V^{l_k}} {g_A^{l_k}} 
\Bigg)
\ee
\nin
where $r_{ij} = \frac{m_i^2}{m_j^2},$ and $\mathcal{F}_1 = 
{g_A^{l_i l_j}}^2 + {g_V^{l_il_j}}^2 ,\;
\mathcal{F}_2 ={g_V^{l_il_j}}{g_A^{l_il_j}}$ function 
contain the FC-parameters. This result shows the 
symmetry-conserving under $g_V^{}$ and $g_A^{},$
and the dependence with the model parameters
($g'_1, M_Z'$ and $\theta'$), explicitly.

The next two subsections, 
we will explore the lepton decays to three 
lepton in the final states. The third subsection 
explores pair of pions in the final state and 
some parameter space in hadron process.

\subsection{\label{ssec:leptons}
$\tau^- \to \mu^- \mu^+ \mu^-$ decay
}
\begin{figure}[H]\centering
\includegraphics[scale = 0.32
]{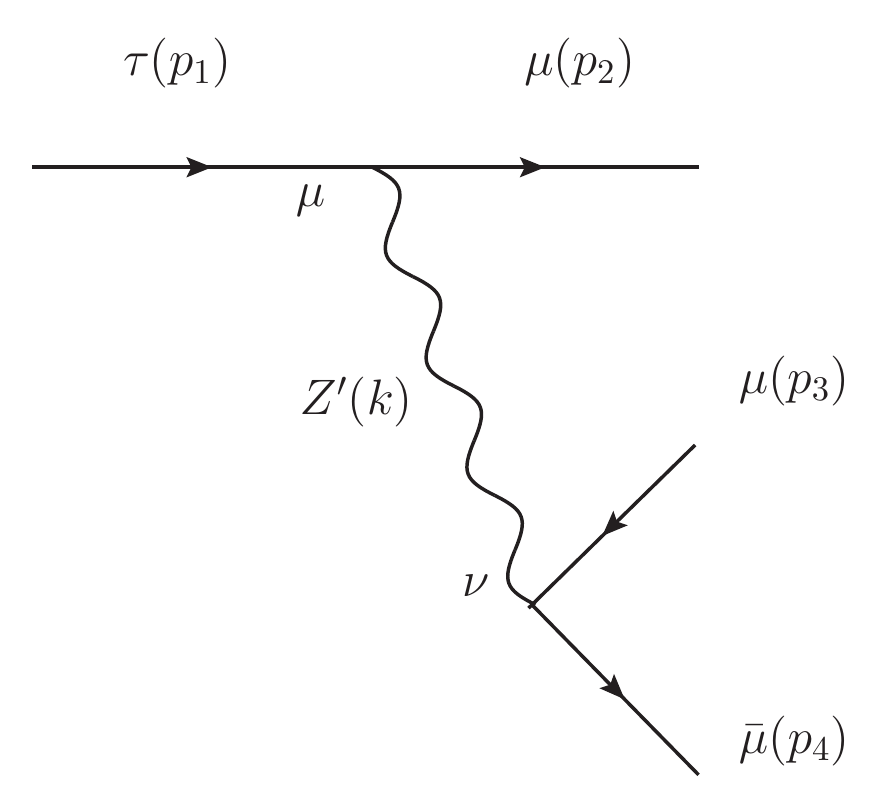}
\includegraphics[scale = 0.32
]{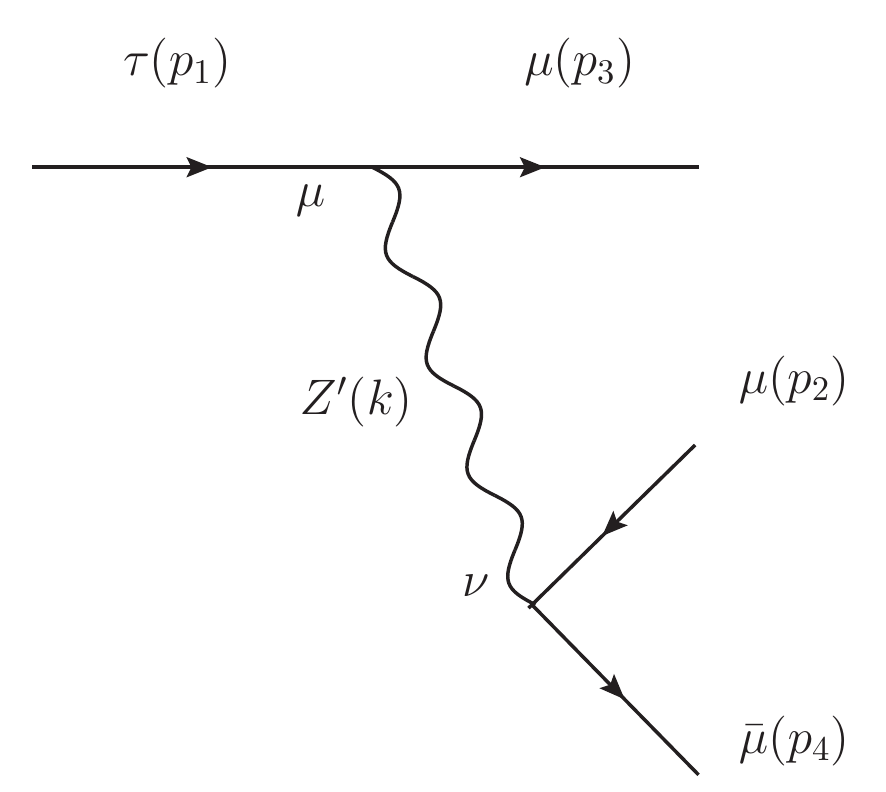}
\caption{\label{fd:tau-mumumu}
Feynman diagram and its adjoint for muons in the final state.
}
\end{figure}

We will explore the process 
represented in fig. \ref{fd:tau-mumumu}. 

The decay width  
is given by eq. \eqref{eq:W_l-lllb}, considering 
 $l_i = \tau^-$ and $l_j = l_k=\mu^-.$ We shall explore 
 some cases for the parameters. We include the 
 experimental data and some phenomenological results 
 in models with new neutral gauge boson.
 
We have explored two cases: I) $g'_{1} = g_{EW}.$ 
In this case we obtain constraints 
on $Z'$ mass for different models  
(see fig. \ref{fig-tau-Zp-mumumu-I}).
Figure \ref{fig-tau-Zp-mumumu-I} shows the 
$\Br(\tau^- \to \mu^- \mu^+ \mu^-),$ we have plotted 
331 models, E$_6$ models 
${Z'}_\chi$, ${Z'}_\psi$ and ${Z'}_\eta$ with 
 $\alpha= 0, \pi/2, \arctan(-\sqrt{5/3}),$
respectively; where
$\alpha-$parameter is an angle to  
define the symmetry breaking pattern of the E$_6$ 
models. 
Left-Right (LR) and alternative Left-Right (ALR) models;
and the horizontal line represents the experimental bound.

The recent results for the 
$\Br(\tau^- \to \mu^- \mu^+ \mu^-) = 2.1\times 10^{-8}$ 
constrains the ${\Br}'$s for the models; 
then we calculated the limits for the $M_{Z'}^{}$ in  
several models those values 
are shown in table \ref{ta:Br-tau-mumumu-I}.
We consider the experimental bounds and 
extract the lower masses for the new neutral 
gauge boson, and obtain limits for this parameter.
\begin{widetext}
\begin{center}
\begin{figure}[H] \centering
\begin{subfigure}{0.48\textwidth}
\includegraphics[scale = 0.95
]{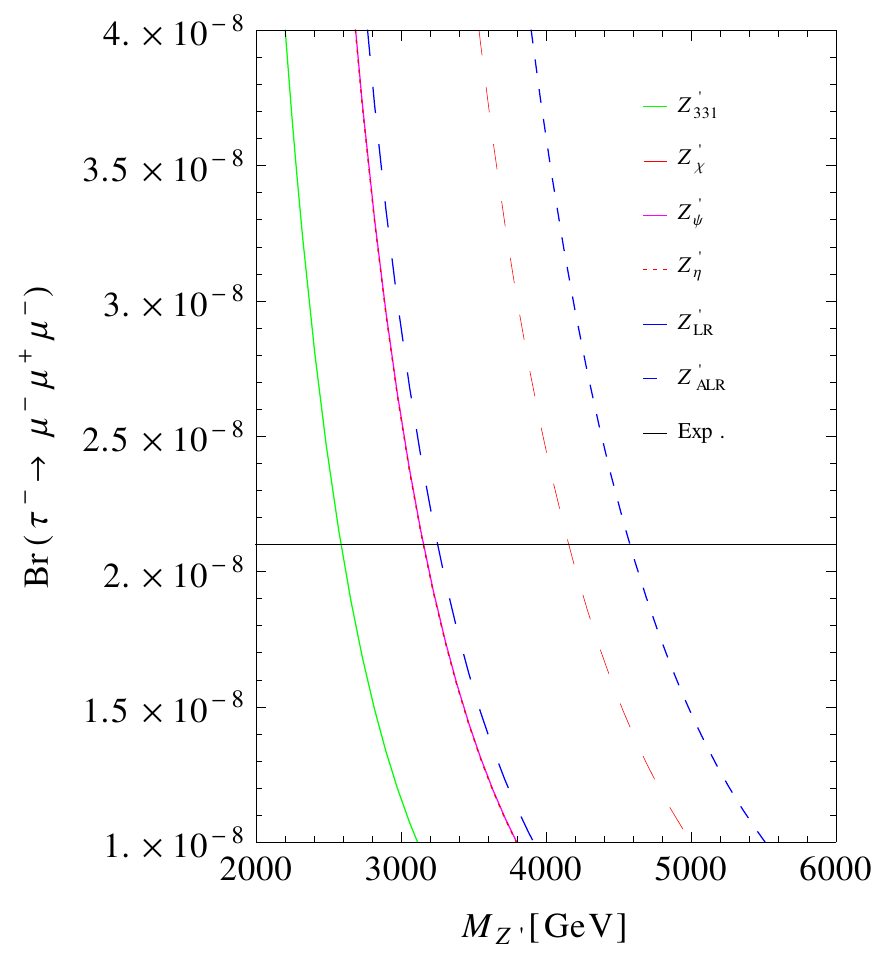}
\caption{\label{fig:fig-mu-Zp-3mu13}
}
 \end{subfigure}
 \begin{subfigure}{0.48\textwidth}
\includegraphics[scale = 0.95
]{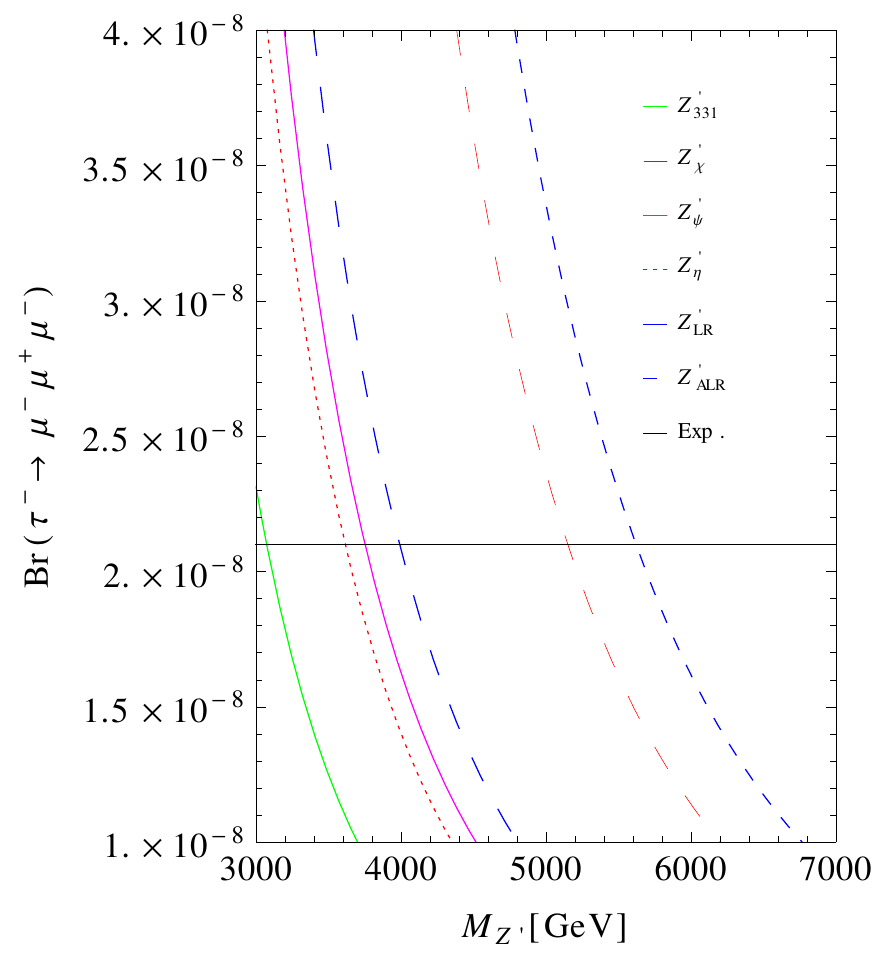}
\caption{\label{fig:fig-mu-Zp-3mu11}
}
 \end{subfigure}
\caption{Branching ratios for the $\tau \to \mu^- \mu^+ \mu^-.$
We chose for a) $g_V^{f_if_j} = 1\times 10^{-1},$
$g_A^{f_if_j} = 1\times 10^{-3}$ and 
b) $g_V^{f_if_j} = 1\times 10^{-1},$
$g_A^{f_if_j} = 1\times 10^{-1}.$ We have taken 
$\theta' = 1\times 10^{-3}$ for each case.
\label{fig-tau-Zp-mumumu-I}
}
\end{figure}
\end{center}
\end{widetext}
We use the previous analysis, and 
found limits over the $Z'$ mass, 
those results can be found in the 
table \ref{ta:Br-tau-mumumu-I}. 
This table shows two scenarios:
first row: we chose an asymmetric  scenario 
where the FC-couplings are different by 
two orders of magnitude. 
Second row: we  chose a symmetric scenario 
where both FC-couplings are same order of 
magnitude. 
Figs. \ref{fig:fig-mu-Zp-3mu13} shows the 
assymetric scenario and 
fig. \ref{fig:fig-mu-Zp-3mu11} shows the 
symmetric scenario, we have explored different 
modes with $Z'$. We found differences in the 
mass scale of the new neutral gauge boson.

\begin{table}[H]
\caption{\label{ta:Br-tau-mumumu-I}
Low allowed mass for $Z'$ in different models. 
The first row shows the mass values for 
fig. \ref{fig:fig-mu-Zp-3mu13}
and the second row shows the mass values 
for fig. \ref{fig:fig-mu-Zp-3mu11}.
We used the couplings given in ref. \cite{Yue:2014yva}.
}
\begin{ruledtabular}
\begin{tabular}{l r r r r r r}
          & $Z'_{331}$&$Z'_{\eta}$&$Z'_{\psi}$&$Z'_{LR}$&$Z'_{\chi}$&$Z'_{ALR}$\\
$M_{Z'}$ (GeV)&$2446.2$&$2984.4$&$3061.4$&$3085.7$&$4025.3$&$4446.1$\\
$M_{Z'}$ (GeV)&$3056.3$&$3622.9$&$3731.4$&$3976.9$&$5160.7$&$5636.3$\\
\end{tabular}
\end{ruledtabular}
\end{table}

For the case II) 
$g'_{1} = 0.105$ taken from \cite{Langacker:2000ju}. 
This case works to constrain every model 
since $M_{Z'} < 900 \GeV$ and this 
mass range is excluded for experiments 
\cite{Beringer:1900zz}.

Next part shows the phenomenology results 
for the leptonic processes mediated by 
a new neutral gauge boson.

\subsection{\label{ssec:leptons}
$\mu^- \to e^- e^+ e^-$ decay
}
\begin{figure}[H]\centering
\includegraphics[scale = 0.32
]{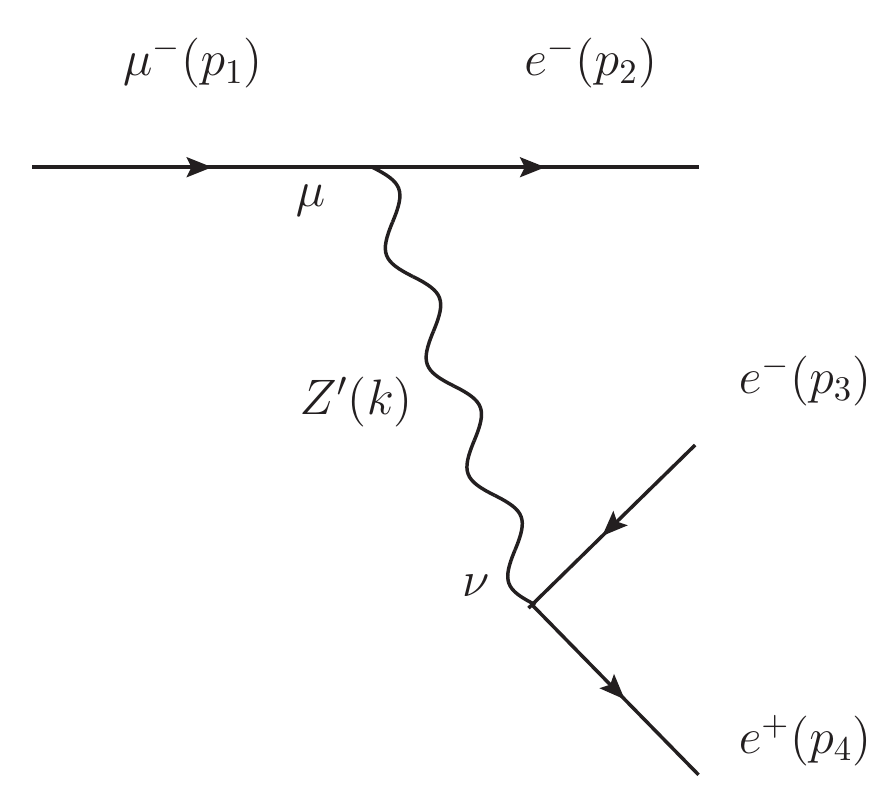}
\includegraphics[scale = 0.32
]{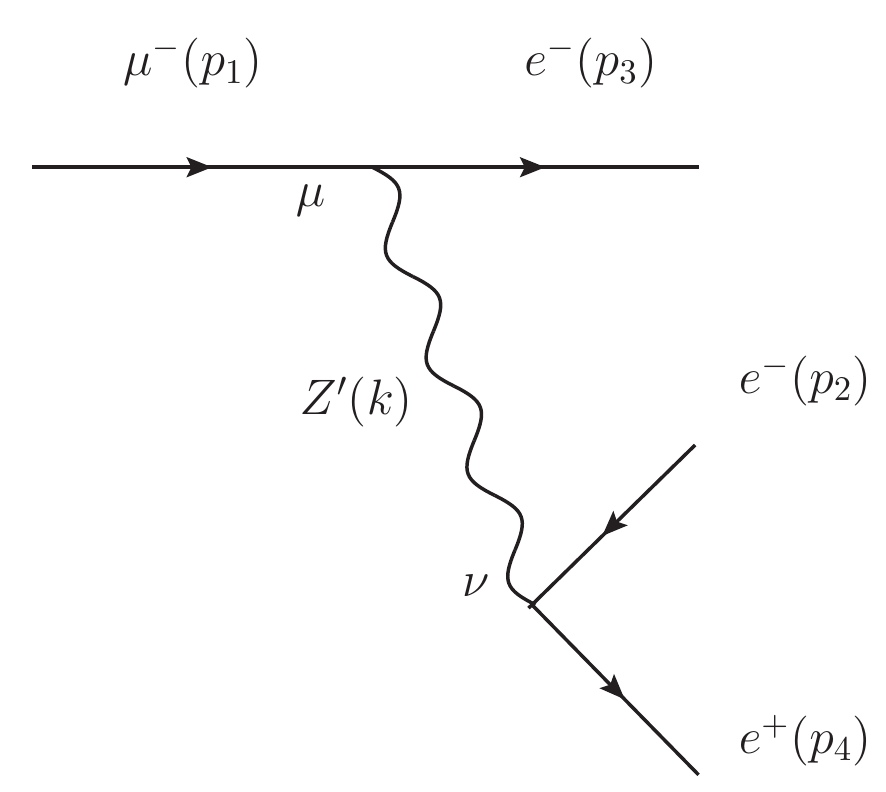}
\caption{\label{fd:tau-eee}
Feynman diagram and its adjoint for electrons in the final state.
}
\end{figure}

\begin{figure}[H]
\centering
\includegraphics[scale = 0.6]{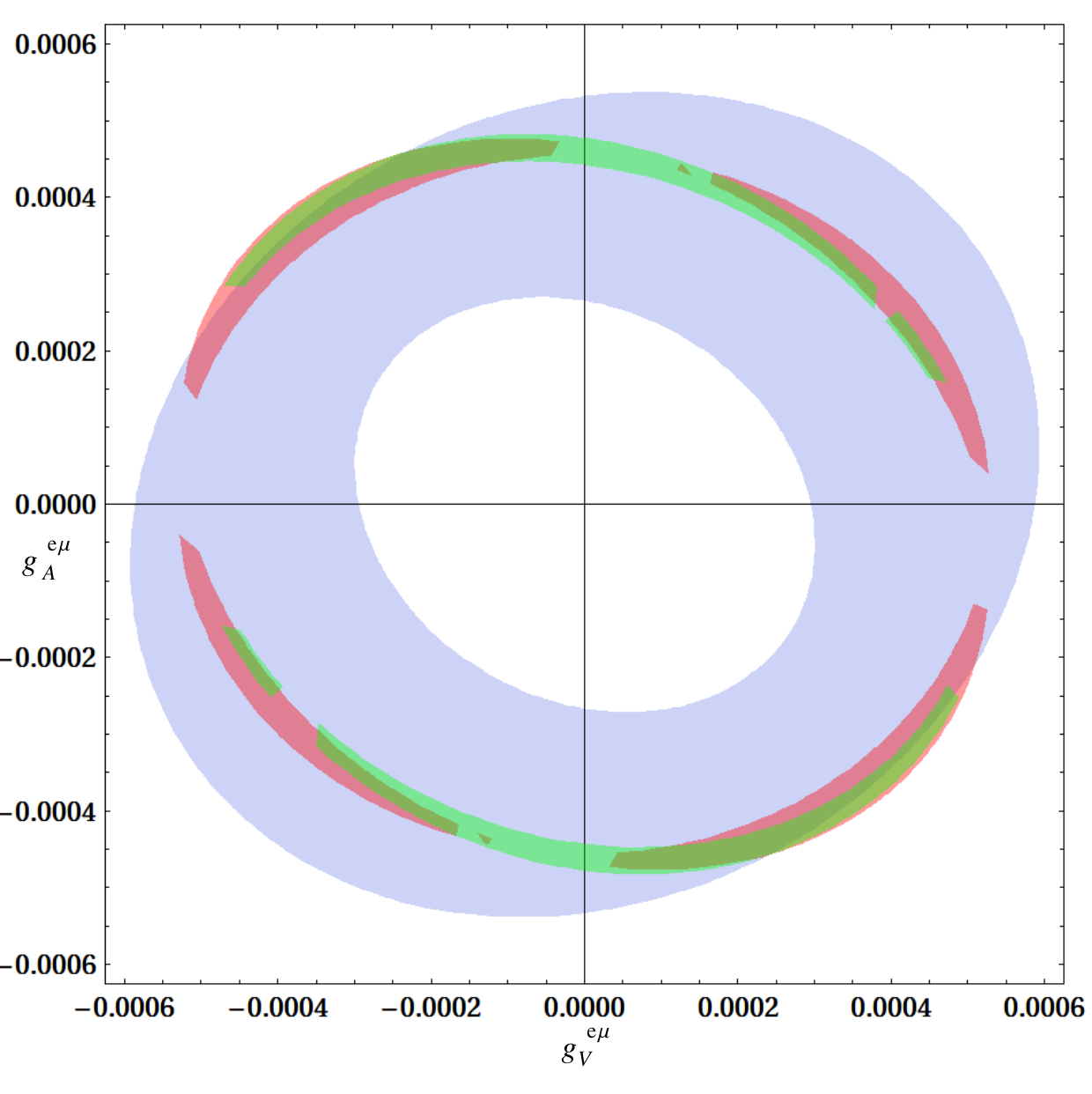}
\caption{$g_V^{f_if_j}$ vs. $g_A^{f_if_j}$ 
using the table \ref{ta:Br-tau-mumumu-I}. Regions among 
curves, which contain the allowed values 
for the $g_{V,A}^{f_if_j}$ parameters. Green region would contain 
the advantaged $g^{e\mu}_{V,A}$ values.
\label{fig-gVfifj-gAfifj-EW}
}
\end{figure}

The Feynman diagram for this process is shown in 
fig. \ref{fd:tau-eee}.
The decay width is given by eq. \eqref{eq:W_l-lllb}, considering
$l_i = \mu^-$ and $l_j = l_k = e^-.$
\nin
From the constraints the last section we explore the 
$\mu^- \to e^-e^+e^-.$ We found $g_{V,A}^{e\mu} \sim {\Or}(10^{-4}),$ it can see 
on fig. \ref{fig-gVfifj-gAfifj-EW} where the green region is favored. 

We imposed the constraints, and 
plotted region, and used the $M_{Z'}$ values 
from the table \ref{ta:Br-tau-mumumu-I}.
In fig. \ref{fig-gVfifj-gAfifj-EW} shows the regions for the different models with a new neutral gauge boson. 
The green region could bound the 
interesting values for the $g_A^{e\mu}-g_V^{e\mu}$ 
parameters.

\subsection{The process in the hadron sector}

We have taken motivations from recent experimental results, which give constraints for the 
hadron and lepton in the final states: 
${\Br}(\tau^- \to \mu^-\pi^+ \pi^-)=
10^{-7}, 10^{-6}, 10^{-5}$
for Belle, BaBar and CLEO, respectively 
\cite{Amhis:2012bh}. 
Even more we expect Belle II has 
surprising physical results when  
it achieves high luminosity (in 2022), 
this is 50 times more than Belle 
\cite{Abe:2010gxa}.  


The hadronic pair $(\pi^+\pi^-)$ is produced, 
at the beginning,  in a initial state 
no hadrons 
$\langle \pi^+(p_4)\pi^-(p_3) 
\big| \bar{Q}\gamma_\mu^{} q\big|0\rangle,$ where 
$Q$ and $q$ are light quarks ($u, d, s$).
The weak current has the form \cite{FloresBaez:2006gf}:

\be
\Big\langle \pi^+(p_4)\pi^-(p_3) 
\Big|
\Big[\frac{b_1 }{2}\Big(\bar{u} \; \gamma^{\mu} \; u - \bar{d} \; 
\gamma^{\mu} \; d\Big)
+
\\\nonumber
\frac{b_2 }{2}\Big(\bar{u} \; \gamma^{\mu} \; u + \bar{d} \; 
\gamma^{\mu} \; d\Big)  
\Big]
\Big|0
\Big\rangle
\ee

\nin
but the second term has not contribution because of the 
G-parity. We have used $ \frac{1}{2}(b_1 + b_2) = a_{uu}$ and 
$ \frac{1}{2}(-b_1 + b_2) = a_{dd}$, with $a_{uu,dd}$ are 
couplings asociated to each state $u\bar{u}, d\bar{d}$. 
Then the hadron element matrix is given by:
\be
\resizebox{.49\textwidth}{!}
{
$
\bigg\langle  \pi^+(p_4)\pi^-(p_3)\bigg| 
\frac{1}{2}(\bar{u} \gamma^{\mu_2} u - \bar{d} 
\gamma^{\mu_2} d)
\bigg|0\bigg\rangle  =  F_\pi(q^2) (p_4 - p_3)^{\mu_2}
$}
\nn
\ee

\nin
where $q^2 = (p_4 + p_3)^2,\;
F_\pi(q^2) = \frac{m_\rho^2}{m_\rho^2 - q^2 - im_\rho \Gamma_\rho}$ and 
$m_\rho = 775 \MeV,\; \Gamma_\rho = 150 \MeV$ \cite{PDG-2014}.
The form factor,  $F_\pi(q^2)$, is the most simple, and 
useful form to use in our model, other forms can be found 
in refs. \cite{FloresTlalpa:2005fz, Diehl:1999ek, Polyakov:1998ze, Roig:2011iv}.  

\begin{figure}[H] \centering
\includegraphics[scale = 0.45]{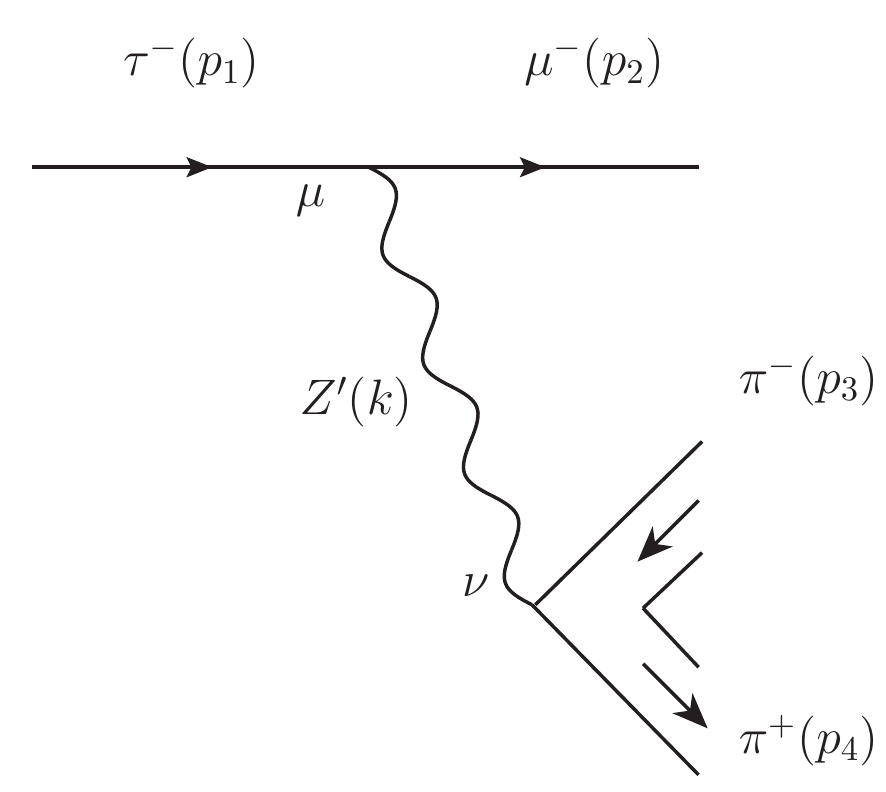}
\caption{The Feynman diagram for the 
$\tau^- \to \mu^- \pi^+ \pi^-$ process.
}
\label{fd:l-lbpipi}
\end{figure}

The Feynman diagram for the process $\tau^- \to \mu^- \pi^+ \pi^-$ 
is shown by fig. \ref{fd:l-lbpipi}. The total amplitud is given by,
\be
{\mathcal{M}}  = F_\pi(q^2)
\, (p_4 - p_3)^{\mu_2} \; 
\Pi^{\mu_1 \mu_2}_{Z'}
\bar{u}(p_2)
g_{Z' \tau\mu}^{\mu_1}
u(p_1)
\ee

\noindent
where $\Pi^{\mu_1 \mu_2}_{Z'} =
{-g^{\mu_1\mu_2}}/{\big(k^2 - M_{Z'}^2\big)}$ and 
$g_{Z' \tau\mu}^{\mu_1}
=
\gamma^{\mu_1} 
\Big({g'}_{V}^{f_if_j} - {g'}_{A}^{f_if_j}\gamma^5\Big).$ 
In this letter we will use $k^2_{} \ll M_{Z'}^{2}.$
Taking $r_{\pi\tau}= \frac{m_{\pi^\pm_{}}^2}{m_\tau^2} \to 0$ 
and $r_{\mu\tau}^{} \to 0,$ 
we obtained the eq. \eqref{eq:dw-tau-2body} for the 
decay width; this is,
\be\label{eq:dw-tau-2body}
\frac{d\Gamma(\tau^{-}\to \mu^- \pi^+\pi^-)}{dxdy}&=&
\frac{m_\tau^{}}{128 \pi^3}
\frac{{r_{\tau Z'}^{2}} \;
r_{\rho\tau}^{2} \;
\mathcal{F}_{VA}^{ij}
\mathcal{G}
}
{ 
\mathcal{H}
}
\ee

\begin{figure}[H] \centering
\includegraphics[
scale = 0.85]{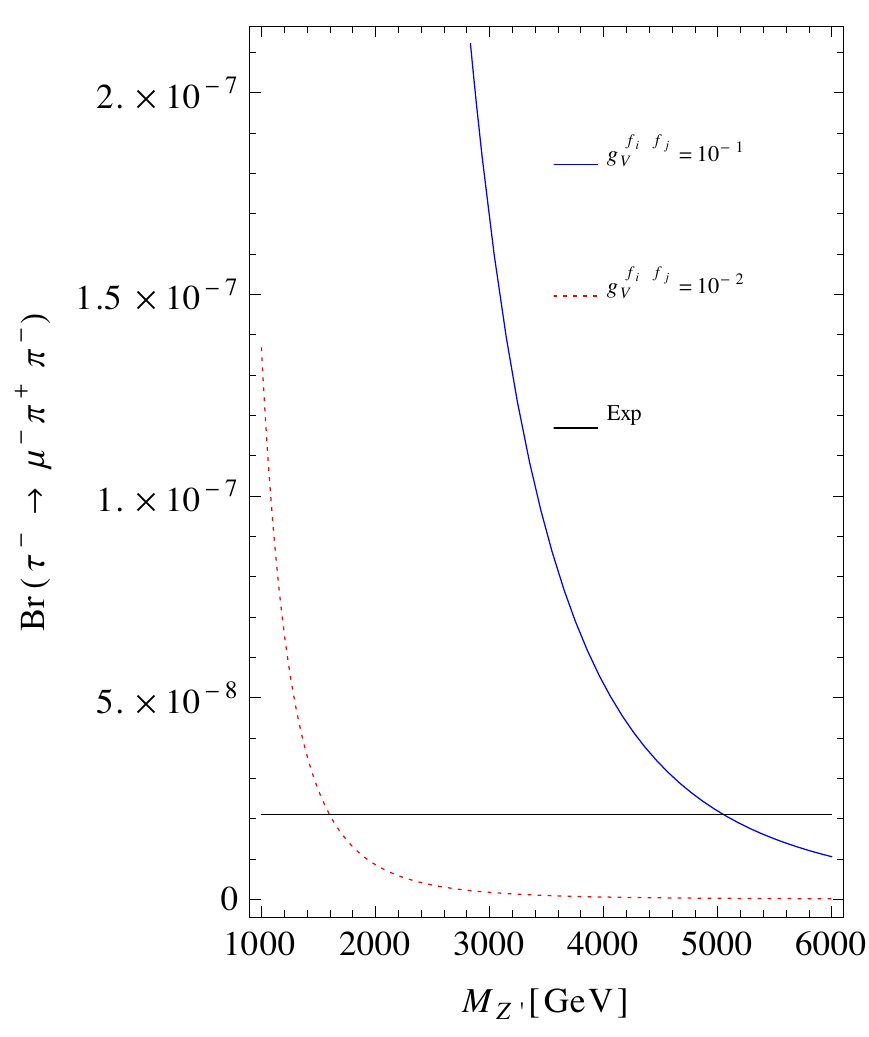}
\caption{Branching ratio for the processes 
$\tau^- \to \mu^- \pi^+ \pi^-$. We show an optimistic  ${g_{V}^{f_if_j}}$ value, 
and we have chosen ${g_{A}^{f_if_j}}=1\times10^{-4}$ for both of them.
}
\label{fig:l-lbpipi}
\end{figure}

\nin
where 
$\mathcal{F}_{VA}^{ij} = 
{g_V^{f_if_j}}^2+{g_A^{f_if_j}}^2,\;$
depends on the FC-parameters;
and $
\mathcal{G}=
x^2+3 x (y-1)+4 (y-1)^2,\;
$ 
$
\mathcal{H} = {r_\rho^{}}^2+\Big({\frac{\Gamma_\rho^2}{m_\tau^2}}+2 y-2\Big)r_{\rho\tau}^{} +(y-1)^2\;
$
contains the variables of integration;  
and $ r_{ij}^{}=\frac{m_i^2}{m_j^2}.$ 
The $x$ and $y$  are variables:
$
1-x <  y  < 1+ x,\;
0 <  x  < 1.
$
\nin
We use the constraints 
given in \cite{Beringer:1900zz} and the results 
is shown in fig. \ref{fig:l-lbpipi}. 
In this process we found that the lowest mass 
is $M_{Z'} \sim 1600 \GeV$, considering the 
lepton flavor-changing, $\tau \to\mu$ and  
some representative values for the couplings, 
since eq. \eqref{eq:dw-tau-2body} depends on
${g_{V,A}^{f_if_j}}$ parameters.

\begin{figure}[H]\centering
\includegraphics[scale = 0.5]{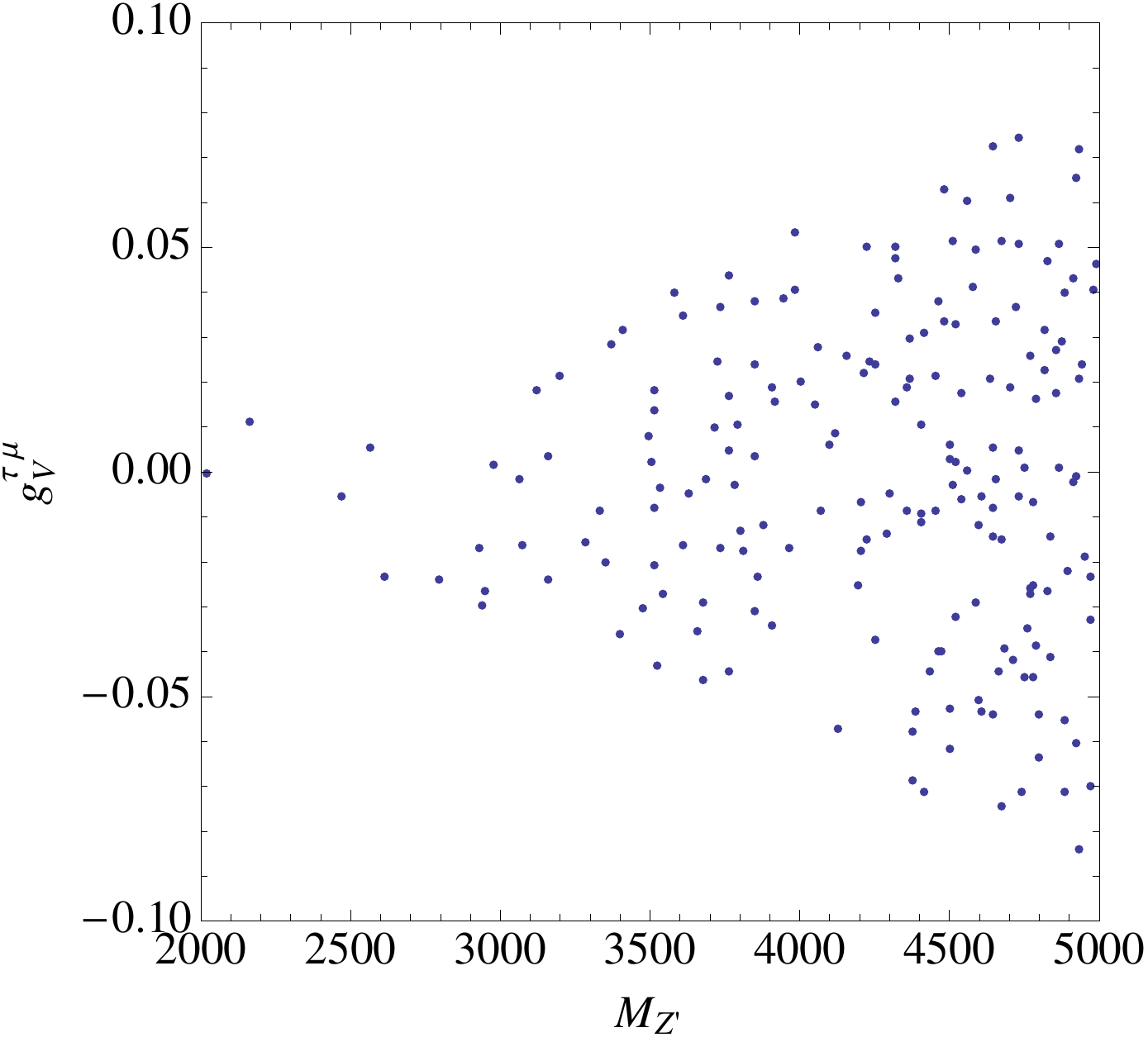}
\caption{\label{fig:gVtmu-MZp}
Scattering plot for the FC-parameters, $M_{Z'}^{},$ 
and ${\Br}$ for the $\tau^-\to \mu^-\pi^+\pi^-$ process.
}
\end{figure}

We explore the parameter space in the 
$\tau^- \to \mu^-\pi^-\pi^+$ process
mediated by a neutral gauge boson. 
We obtained the region for $M_{Z'}^{}$ 
in fig. \ref{fig:gVtmu-MZp}.
This figure shows the allowed region for the 
vector coupling considering the flavor-changing 
mediated by a $Z'$ coming from different models; 
our results show wider region for high $Z'$ 
masses. We expect interesting results for the 
next generation of colliders.

We considered the $g_A^{\tau \mu}$-parameter 
versus $M_{Z'}^{}$ and the result was very similar
what was shown in fig. \ref{fig:gVtmu-MZp}.

\section{\label{sec:discussion}
Discussion and conclusions}

In this letter we explore some models with a $Z'$ which 
has family non-universal  coupling in 
processes with FC as 
$\tau^- \to \mu^- \mu^+ \mu^-$ and $\mu^- \to e^- e^+ e^-,$
and the $\tau \to \mu^- \pi^+\pi^-$ process 
(hadronic process). 
We have obtained estimation for the $Z'$ 
mass and its parameter,  
and considering the current experimental 
results, we found some representative values
 for the $M_{Z'}$ and regions for the parameters, 
which contains the FNU charges. I am aware that this is a simple method to get 
bounds, however it works to give a nominal 
values for the $Z'$ mass which obey the 
experimental constraints.

Figure \ref{fig-tau-Zp-mumumu-I} ~shows a 
excluded mass range for different models in 
lepton processes considering FC, 
this mass range could be explored for next colliders. 
From the $\tau^- \to \mu^- \mu^+ \mu^-$ process, 
we report $M_{Z'} \gtrsim 2500 \GeV$ for 
$g'_1=g_{EW}$ (gauge coupling) and 
$g_V^{f_if_J} = 10^{-1}, g_A^{f_if_J} = 10^{-3}$ 
or viceversa, since equations are 
symmetrical under $g_V^{f_if_J} \leftrightarrow g_A^{f_if_J}.$
For the process $\mu^- \to e^- e^+ e^-,$  
we note $M_{Z'} \gtrsim 3000 \GeV$ 
for $g_{V,A}^{f_if_J} \lesssim 10^{-4};$ 
and from $\tau^- \to \mu^- \pi^+ \pi^-$ process,
we found $M_{Z'} \gtrsim 5000 \GeV,$
considering $g_V^{f_if_J} = 10^{-1,-2}$ and  $g_A^{f_if_J} = 10^{-4}.$

We regard the experimental bounds for the 
$l_i^{} \to l_i^{} l_j^{}\bar{l}_k^{}$ process, 
which restrict the parameter space 
(see fig. \ref{fig-gVfifj-gAfifj-EW}): the  bounded region  
for the overlapping gives the most interesting values for 
the vector and vector-axial couplings, considering FC in 
lepton process. Exploring the hadron processes, we 
also exclude a mass range, figure \ref{fig:l-lbpipi}.  
The excluded regions are consistent with the experimental 
results \cite{Agashe:2014kda}, 
see table \ref{ta:Br-tau-mumumu-I}.

Figure \ref{fig:gVtmu-MZp} shows the allowed region 
considering the current experimental results for the 
$\tau^-\to\mu^-\pi^+\pi^-$ process. We found a similar 
region for the $g^{\tau\mu}_{A}-$parameter. In general, 
this plot reveals the $M_{Z'}$ values for the coupling 
with FC. In this escenarios with new physics, 
correlation between those parameters 
could give more information about the 
FC in the hadron sector.

Using the recent results from LHC, 
we have constrained the 
$g_{V,A}^{f_if_j}$ parameters, besides found the lowest mass 
allowed for a $Z'$ coming from 
some models, considering 
flavor-changing neutral currents, family 
non-universal coupling; and a process comes from hadron sector.
Though we report high mass limits for a new neutral gauge 
boson, we are optimistic about the next LHC results as well as the future generation colliders.

\begin{acknowledgments}
This work was supported by postdoctoral scholarships at 
DGAPA-UNAM and SNI (Conacyt). I thank G. 
L\'opez Castro for the discussions about physics; 
I would also like to thank P. Roig, M.A. P\'erez-Ang\'on and R.Gait\'an Lozano
for reading this manuscript, and A. Courtoy for providing comments 
and refs. \cite{Diehl:1999ek, Polyakov:1998ze, Altmannshofer:2016brv}.
\end{acknowledgments}

{\bf{Final remarks:}} 
I want to share several preprints on FC and new boson phenomenology uploaded while 
I was writing this manuscritpt:  
\cite{Khachatryan:2015kon, Fuentes-Martin:2015owx, Banerjee:2015hoa,
  Boyarkin:2016fah, Gutierrez-Rodriguez:2015qka, Kim:2016bdu}.

	  \bibliography{biblio-Zp-LF}
\end{document}